\documentclass[journal]{osa-article}
\usepackage{siunitx}  
\usepackage{graphicx}
\usepackage[numbers]{natbib}

\journal{osajournal}
\articletype{Research Article}
\begin{document}

\title{500~W Peak Power Cavity Dumped 2-micron GaSb-Based VECSEL}
\author{Jacob~Hoehler\authormark{1,2},
        Ricky~Gibson\authormark{2},
        Jennifer~M.~Reed\authormark{2,3},
        and~Robert~Bedford\authormark{2,4,*}
    }

\address{
\authormark{1}University of Dayton, Department of Electro-Optics and Photonics, 300 College Park, OH 45469, USA\\
\authormark{2}Air Force Research Laboratory, Sensors Directorate, Wright-Patterson Air Force Base, OH 45433, USA\\
\authormark{3}KBR Laboratories, Beavercreek, OH 45431, USA\\
\authormark{4}Air Force Research Laboratory, Materials and Manufacturing Directorate, Wright-Patterson Air Force Base, OH 45433, USA
}

\email{\authormark{*}robert.bedford@us.af.mil} 

\begin{abstract}
A single transverse mode high pulse-energy VECSEL was developed.
The GaSb based VECSEL emits at a wavelength of  2.04 microns with a peak power exceeding 500~W while maintaining good beam quality.
The cavity employs a Pockels cell combined with a low-loss thin film polarizer to selectively dump the intracavity energy into a 10~ns pulse. The laser has promise for incoherent LiDAR, materials processing, gas sensing, and nonlinear optics.
\end{abstract}

\section{Introduction}\label{sec:1}

Vertical-external-cavity surface-emitting-lasers (VECSELs) utilize a semiconductor gain medium which offers a versatile platform, offering spectral ranges from visible to the mid-wave infrared, as well as temporal output, from high-repetition rate femtosecond pulse trains to continuous wave (CW) output, owed to the semiconductor gain medium \cite{okhotnikov_semiconductor_2010,guina_optically_2017}. In addition, the external cavity enables wider spectral access by allowing for the insertion of intracavity nonlinear frequency conversion crystals which extend the output wavelengths from the UV-C \cite{8084098} and into the terahertz regimes \cite{bondaz2019generation}. In the two micron spectral wavelength range GaSb VECSELs have demonstrated $>$20~W CW output power \cite{holl_power_scaling_GaSb} achieved at a gain chip temperature of \(-3~\si{\celsius}\),  and \(70~\si\watt\) peak power for a gain-switched \(200~\si{\nano\second}\) pulse \cite{yarborough2009record}.  Gain-switching can be used to access pulse-lengths between roughly the photon lifetime ($\sim$100~ns) and quasi-continuous wave operation.  Conversely, mode-locked operation can be achieved through the use of nonlinear elements such as saturable absorber mirrors (SESAMs) to facilitate access to the picosecond \cite{harkonen2010picosecond} and femtosecond \cite{harkonen2011modelocked} regimes, dictated by the nonlinear dynamics of the gain and mode-locking element.  In addition to mode-locking two micron VECSELs, GaSb based SESAMs have been used to mode-lock rare earth \cite{zhao2020ceramic} and transition metal \cite{barh2021znsOscillator} lasers as well as VECSELs in the near-infrared \cite{Harkonen2018gasbSesam}. Further details on GaSb SESAMs characterized from the short-wave infrared (SWIR) to the mid-wave infrared (MWIR) can be found in \cite{heidrich2020sesam}.

Between the gain-switched and mode-locking regimes, the few nanosecond regime can also be reached with an intracavity fast-switching element such as the Pockels cell \cite{kaspar_electro-optically_2012}.
The nanosecond pulse configuration is a form of $Q$-switching known as cavity dumping. Here, a high-$Q$ cavity is utilized to store energy rather than within the gain medium.  This concept was first demonstrated in a Nd:YAG laser showing that an energetic pulse was limited only by the cavity length of the laser \cite{hook1966laser}, where pulse lengths are generally dictated by the round trip time of a photon within the cavity ($2L_{c}/c$, where $L_{c}$ is  the cavity length). In order to facilitate cavity dumping, a fast intracavity element must be used to rapidly dump the stored energy within the cavity. As an example, an acoustic optic modulator (AOM) can be used to deflect the output beam; alternatively, a Pockels cell inducing a rapid polarization change in combination with a polarization selective output coupler is an effective fast dumping element. Because the typical semiconductor carrier lifetime doesn't permit significant energy storage within the gain, the cavity dumping approach is required for VECSELs to achieve energetic pulses.  This is contrasted with solid state lasers where traditional $Q$-switching is often utilized. Such nanosecond pulses achievable via cavity dumping are fitting for long-range LiDAR to circumvent some deficiencies seen in shorter pulses \cite{melngailis1996laser}. While $Q$-switched rare earth and fiber lasers have long led the field for LiDAR sources, typical emissions are below the eye-safe range of \(1.5~\si{\micro\meter}\), and suffer from increased manufacturing time and costs \cite{mcmanamon_lidar_2019}.  The use of a 2040~nm GaSb laser places it between the typical spectral emission of holmium and thulium as well as within a narrow atmospheric transmission window. Lasers operating in the two micron spectral range also have applications in gas sensing, optical communication, and in the medical field \cite{scholle20102}.

Cavity dumped VECSELs at $\lambda$ $\sim$ \(1~\si{\micro\meter}\) have been demonstrated with peak powers of \(57~\si\watt\) over a \(30~\si{\nano\second}\) pulse utilizing an AOM \cite{savitski_cavity-dumping_2010}. In this case, the pulse length was longer than the photon cavity round-trip and the temporal behavior was dominated by the slow transition time of the AOM. Conversely, a Pockels cell can be designed to switch more rapidly. Additionally, by incorporating gain-switching  to mitigate thermal effects, higher peak powers and sharper pulses have been achieved reaching \(1~\si{\kilo\watt}\) for a single-gain-chip VECSEL \cite{myers_high_2017} and \(1.7~\si{\kilo\watt}\) for a dual-gain-chip VECSEL \cite{myers2017single} for \(3-5~\si{\nano\second}\) pulses. In the $\lambda$ $\sim$ \(2~\si{\micro\meter}\) regime, cavity dumped VECSELs have reached \(30~\si\watt\) in a \(3-\si{\nano\second}\) pulse \cite{kaspar_electro-optically_2012}. In this manuscript, we report high peak power of a cavity dumped $\sim2~\si{\micro\meter}$ VECSEL by designing a low-loss cavity, and incorporating gain switching for thermal mitigation.

\section{Experimental Setup}\label{sec:2}

We start with a GaSb-based gain-chip designed to emit near 2040~nm where a narrow atmospheric transmission window exists. This nine-quantum-well InGaSb/GaSb gain structure was grown on a high-reflectivity GaSb/AlAsSb distributed Bragg reflector (DBR) and was capillary-bonded to a single-crystal wedged diamond heat spreader on the surface to efficiently remove heat with very little optical loss.  In addition, the substrate of the chip was thermally contacted to the submount \cite{PrivateDiscussionWithVEXLUM}. The subassembly was then mounted onto a water-cooled copper block and maintained at \(16~\si{\celsius}\).

Based on simulations outlined in Ref~\citenum{bedford_large_2018}, and using the cavity geometry detailed below, we determined the ideal cavity/pulse length to fully deplete the cavity in a single round trip required the cavity round-trip time to be $\geq$3$\times$ the Pockels cell switching time.  The Pockels cell used had a manufacturer-quoted switching time of  \(3~\si{\nano\second}\) and therefore  a round trip time of  \(10~\si{\nano\second}\) was suitable, leading to a cavity length  of \(1.5~\si\meter\).

\begin{figure}[t!]
  \centering
  \fbox{\includegraphics[width=\linewidth]{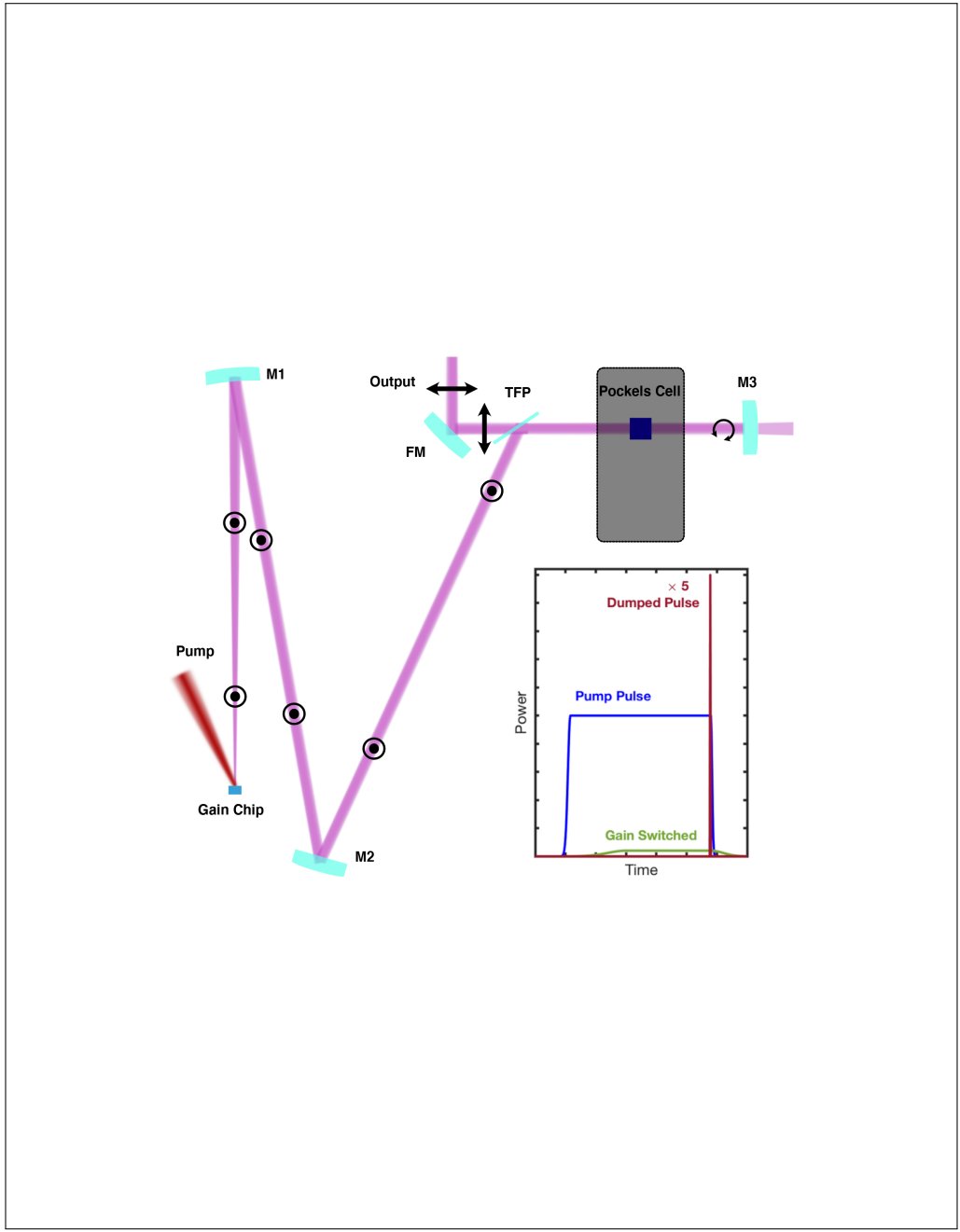}}
  \caption{A schematic of the cavity utilized for cavity dumping with polarization symbols overlaying the cavity mode. The overall cavity length is \(1.5~\si\meter\) and features four legs defined by the high-reflectivity cavity mirrors ($M_1$,$M_2$, and $M_3$) and thin film polarizer (TFP) along with a Pockels cell and a fold mirror (FM) external to the cavity. INSET: A comparison of the temporal and peak power characteristics of the idealized pump pulse (blue), gain switched pulse leaking out of the TFP (green), and dumped pulse (maroon, amplitude reduced by a factor of 5).}
  \label{fig.cdCavity}
\end{figure}

The cavity geometry is displayed in Fig.~\ref{fig.cdCavity}.
The gain chip composes one end of the cavity and was pumped using a $\lambda\approx$ \(980~\si{\nano\meter}\)  fiber-coupled diode laser bar. The \(200~\si{\micro\meter}\) diameter core fiber was magnified 1.8$\times$ for an on-chip spot size of \(364~\si{\micro\meter}\) at a angle of  20~degrees. This pump size matches the mode size well, with a spot-size to pump-size ratio of \(\approx 0.96\), limiting the number of higher-order spatial modes which may be able to lase \cite{laurain_modeling_2019}.

The first mirror, \(M_1\), in this cavity featured a reflectivity of \(99.8\%\) and a \(600~\si{\milli\meter}\) radius of curvature. In order to provide a large spot size on the chip and (roughly) collimate the mode throughout the rest of the cavity, the length of the first leg was set at \(375~\si{\milli\meter}\) with a five degree angle of incidence minimizing astigmatism, while allowing the beam to clear the gain chip's mount in the following leg. A flat mirror, \(M_2\) with reflectivity greater than \(99.9\%\) was then placed \(455~\si{\milli\meter}\) from \(M_1\)  to comprise leg 2 and accomodate a smaller cavity footprint. This mirror was oriented  at a ${16~\si\degree}$ angle of incidence, ultimately allowing access to the output port from the thin-film polarizer (TFP).

A TFP with \(99.7\%\) reflectivity for \(S\)-polarized ($>$\(99\%\) transmission for p-polarized) was selected as the polarization-selecting optic such that with the Pockels cell, fast adjustment of the cavity-$Q$ is facilitated. The TFP was placed \(498~\si{\milli\meter}\) from $M_2$ and positioned at Brewster's angle such that the reflective state completed the resonator while the transmissive state would act as the dumped output achieving the greatest efficiency. This forces the cavity to lase in \(S\)-polarization and produces lower losses when compared to utilizing the transmissive state of the TFP. The end mirror, \(M_3\), identical to \(M_2\), was placed \(208~\si{\milli\meter}\) from the TFP.

Two considerations lead to the placement of the Pockels cell. First, we chose to utilize a $\lambda$/4 pulse as in Ref~\citenum{myers2017single} for optimizing the switching time; this necessitates minimizing the cavity length between the end mirror and the Pockels cell, eliminating any side pulses resulting from  partial polarization rotation \cite{bedford_large_2018}. Secondly, maximizing the beam size in the Pockels cell minimizes the optical power density in the nonlinear crystal. This, in turn, increases the possible intracavity power before detrimentally reducing the $Q$ of the cavity. Therefore, the Pockels cell, constructed of a rubidum titanyle phosphate (RTP) nonlinear crystal, was placed in the final cavity leg between the polarizer and end mirror. Both of the crystal facets and the enclosing windows are AR coated for nominally \(\lambda\sim 2~\si{\micro\meter}\). The Pockels cell is driven by a high voltage DC power supply which supplies a bias of \(\approx700~\si{\volt}\) and for a single pass quarter wave rotation a \(1900~\si{\volt}\) pulse. The Pockels cell driver is limited to pulse repetition frequency (PRF) of \(5~\si{\kilo\hertz}\).

Finally, gain switching of a cavity dumped VECSEL  was previously shown to increase the peak power by roughly an order of magnitude \cite{myers_high_2017} which was attributed to mitigating thermal effects. The same pumping scheme is employed in this work, inset of Fig.~\ref{fig.cdCavity}. The pump diode driver is triggered by a digital delay generator which then also triggers the Pockels cell, thus dumping the pulse \(\approx 23~\si{\micro\second}\) later,  once the intracavity power has plateaued (c.f. Fig.~\ref{fig.undumpTrace}).

\section{Experimental Results}\label{sec:3}

\begin{figure}[h!]
  \centering
  \includegraphics[width=\linewidth]{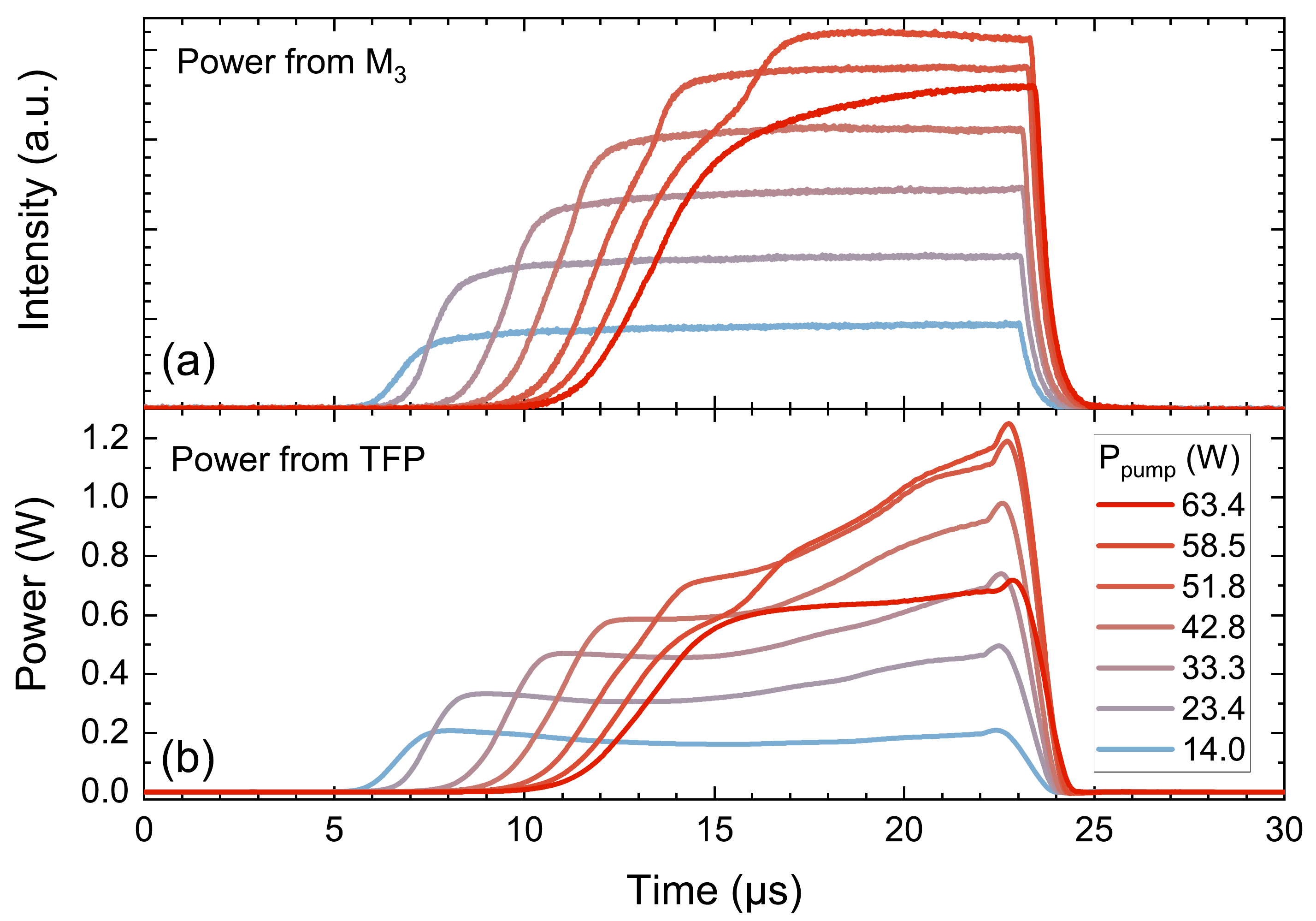}
  \caption{(a) The undumped cavity traces from behind, $WM3$, as measured by a \(25~\si{\mega\hertz}\) photodiode. (b) Undumped traces out leaking out of the thin film polarizer output coupler during the cavity build-up as measured by a \(12.5~\si{\giga\hertz}\) photodiode.}
  \label{fig.undumpTrace}
\end{figure}

In order to trigger the Pockels cell for an optimized peak output power, we must allow for the optical energy to build within the cavity. In order to determine the optimum delay time, the emission leaking out the end mirror, \(M_3\), is collected using a \(25~\si{\mega\hertz}\) extended range InGaAs single channel detector. We allow for the intracavity power to reach its maximum at which point dumping the output power results in the most efficient energy conversion. This effectively optimizes the gain switching configuration by minimizing thermal load. The build-up as measured from \(M_3\) is displayed in Fig.~\ref{fig.undumpTrace} (a) for a pump pulse width of \(23.2~\si{\micro\second}\).

\begin{figure}[h]
  \centering
  \includegraphics[width = \linewidth]{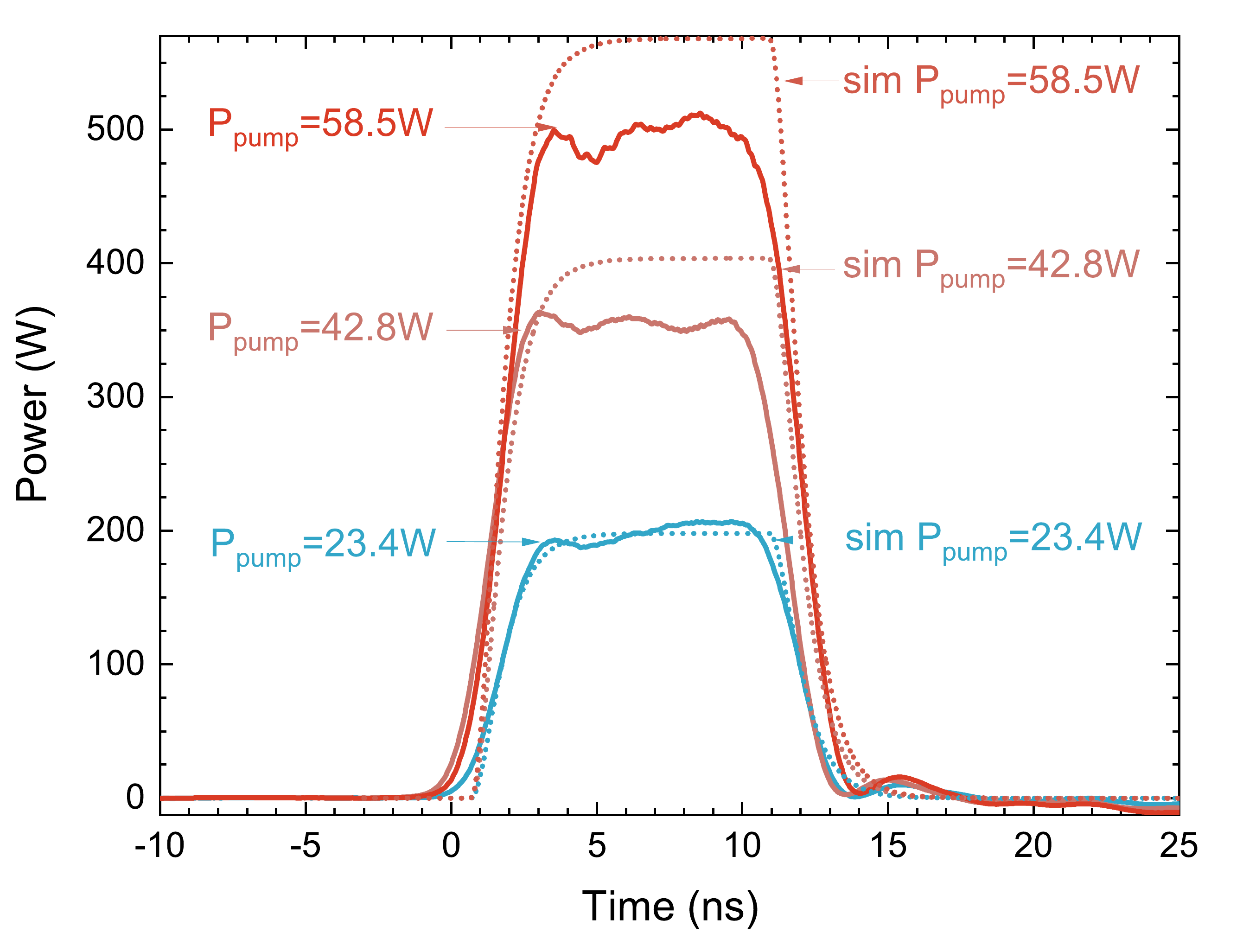}
  \caption{The dumped pulse out of thin film polarizer scaled to the pulse power. The dotted curves are simulated pulses for a  \(1.5~\si{\nano\second}\) Pockels cell switching time at indicated pump powers. t = 0 corresponds to \(22.9~\si{\micro\second}\) after the pump pulse is triggered.}
  \label{fig.dumpTrace}
\end{figure}

The undumped and dumped pulse energies are measured from the emission out of the TFP from the average of 300 pulses using a microJoule thermopile energy meter.  Due to the disparate times scales the gain switched pulse energy leaking out of the TFP and the dumped pulse, the two pulses energies are on the same order of magnitude. This calculation therefore requires collection of the temporal pulses out of the TFP for both the undumped and dumped outputs which are shown in Fig.~\ref{fig.undumpTrace} (b) and Fig.~\ref{fig.dumpTrace}, respectively. Therefore, the low power leakage up to the time when the dumped pulse starts, as seen in the inset of Figure~\ref{fig.cdCavity}, must be subtracted from the dumped output for an accurate measurement of the dumped pulse power.  The integration of the dumped pulse is then normalized to unity and multiplied by the remaining energy. The peak power is then read as the maximum value of the pulse.

The undumped pulse collected similarly to the signal out of $M_3$ and normalized by the measured undumped pulse energy would be expected to match the temporal shape measured out of \(M_3\) in Fig.~\ref{fig.undumpTrace}~(a). The difference between the shapes in Fig.~\ref{fig.undumpTrace}~(a) and (b) was found to be caused by a thermally-induced birefringence within the Pockels cell, even after gain switching was employed. Since a polarization shift would have a greater influence in the loss of a polarization-dependent element, this rotation was not observed from $M_3$.

The dumped pulses have a width of \(\approx10~\si{\nano\second}\), matching the round-trip time of the cavity. The sharp rising/falling edges of the pulse offer an indication that the Pockels cell was switching faster than specified by the vendor; the Pockels cell switching time in this configuration was roughly \(1.5~\si{\nano\second}\). Three simulated pulses are also plotted in Fig.~\ref{fig.dumpTrace} for selected pump powers. The simulation details are given below.

\begin{figure}[h!]
  \centering
  \includegraphics[width = \linewidth]{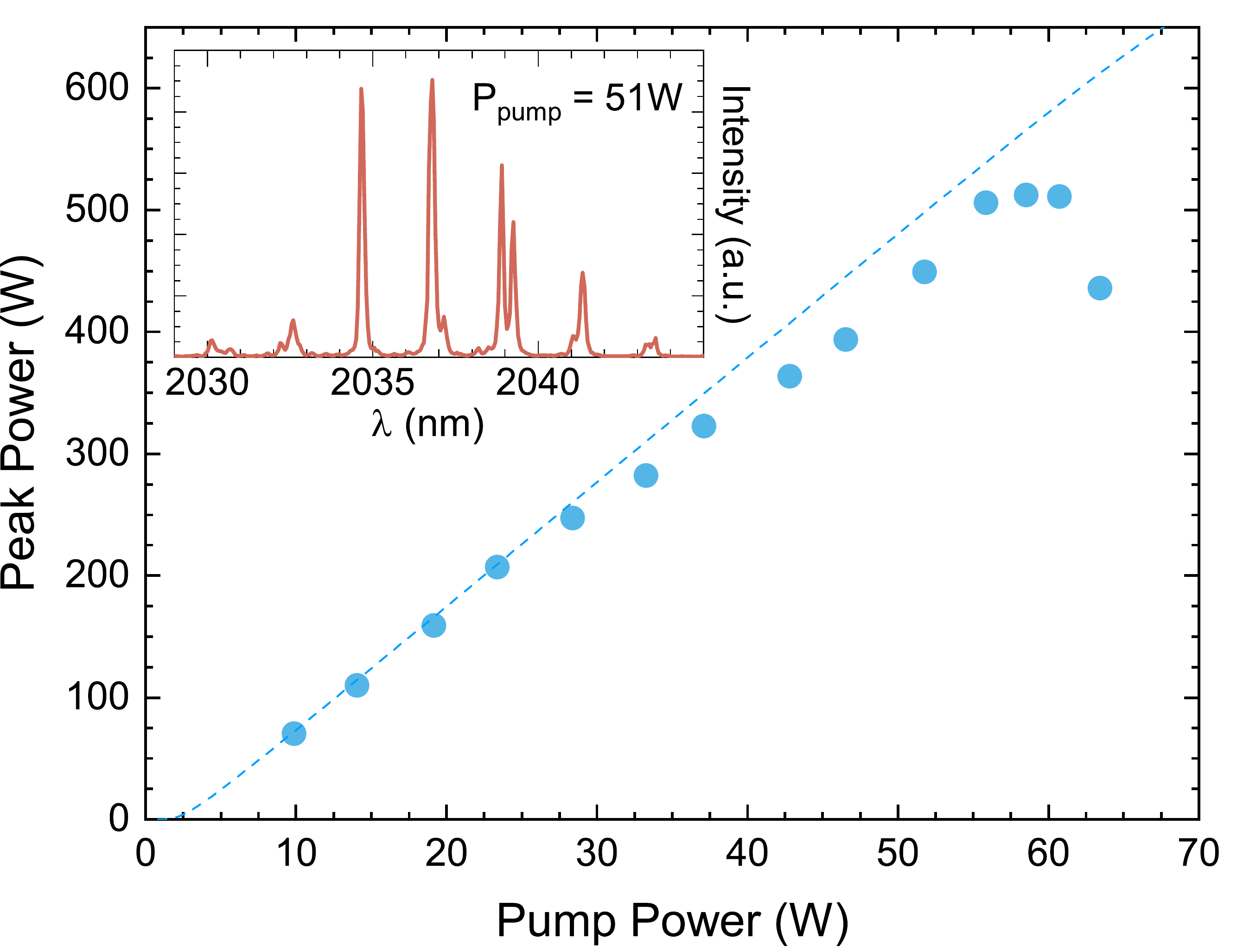}
  \caption{The LI curve as peak power of the cavity dumped VECSEL coupled out of the thin film polarizer in a 10~ns pulse (dashed curve simulated). INSET: Spectrum of the dumped pulse at a pump power of 51W.}
  \label{fig.liCurve}
\end{figure}

The LI curve is shown in Fig.~\ref{fig.liCurve} and represents the resulting peak powers of the dumped pulses in Fig.~\ref{fig.dumpTrace}, both experimental and simulated. A peak power of \(512~\si\watt\) is reached for a \(55~\si\watt\) pump pulse before a sharp rollover above \(60~\si\watt\) of peak pump power. The laser's spectral output from the TFP was measured in various operating states with a spectrometer with a resolution of \(0.25~\si{\nano\meter}\) and a single channel extended range InGaAs detector. Spectral characteristics for the system  below roll-over, 51.2~W pump power, may be seen in the inset of Fig.~\ref{fig.liCurve} and shows a collection of discrete peaks ranging from \(2025-2045~\si{\nano\meter}\). The spacing between and the shape of the peaks result from filtering of various intracavity elements with the greatest influence from the diamond intracavity-heatspreader.  While the location and number of peaks remained consistent across the LI curve, the power distribution of these peaks would also red-shift as the gain chip temperature increased, with the largest peak near \(2037.5~\si{\nano\meter}\) at thermal rollover.

To evaluate beam quality, we utilized an \(M^2\) measurement. The output  from the laser pumped with 51 W power behind $M_3$ was focused using a \(250~\si{\milli\meter}\) focal length lens and measured by a microbolometer beam profiler which was translated along the beam's propagation axis  \(100~\si{\milli\meter}\) to \(450~\si{\milli\meter}\) behind the lens via a \(600~\si{\milli\meter}\) translation stage. 
Due to the low duty cycle, the $1/e^2$ value of the beam radii are used for fitting the \(M^2\) making the background of the microbolometer less influential in the resulting fitting parameters. The spatial profiles were fit to a single Gaussian beam, plotted against the propagation distance, and fit to the Gaussian propagation equation. The results give an \(M^2\)  of 1.003 and 1.020 in the two orthogonal transverse dimensions for all pump powers. These results suggest that the beam is single transverse mode through the entire operation range as expected based on the pump and mode spot sizes on the chip.

\section{Discussion}\label{sec:4}

The \(512~\si{\watt}\) peak power achieved makes this compact laser promising for the applications described previously. Although there are several critical improvements to attain this result, the gain switching employed for this work reduces the thermal load on the chip and has been shown to improve the peak power of GaAs and GaSb based VECSELs \cite{yarborough2009record, burns2009recent, laurain2011high} and in cavity dumped VECSELs at $\lambda\sim$  \(1~\si{\micro\meter}\) by increasing the peak dumped power by a factor of $\sim$10 \cite{myers_high_2017}. In this work, as in previously reported cavity dumped VECSELs, the Pockels cell  is the largest addition to the overall cavity loss, a contribution of \(\approx 2\mbox{--}3 \%\). Even for a 500~W output that would represent an output power loss in the few Watts range.

\begin{figure}[h!]
  \centering
  \includegraphics[width = \linewidth]{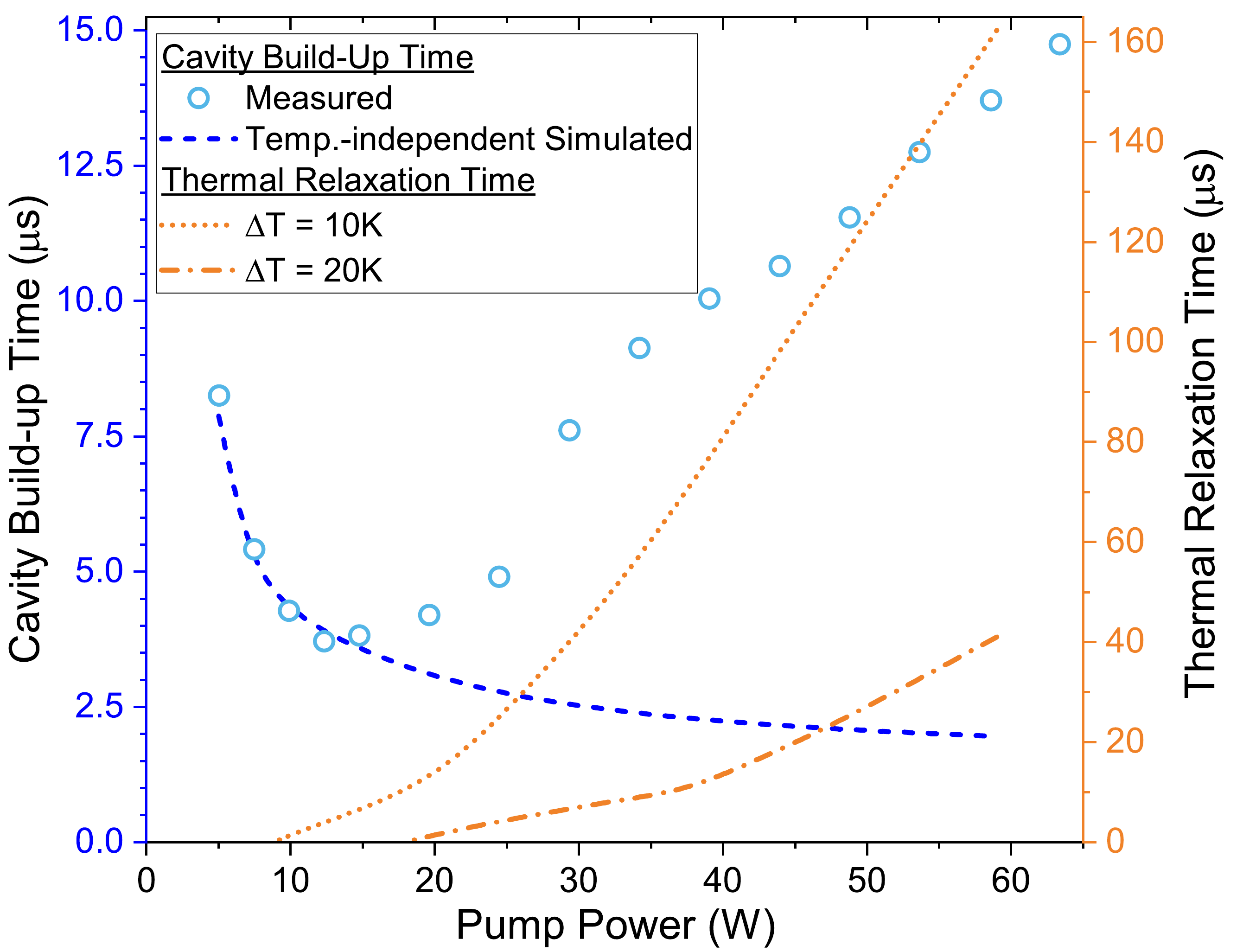}
  \caption{The experimentally measured build-up time (light blue circles), the simulated build-up time based on carrier dynamics (blue dashed line), and the thermal relaxation time from pulsed thermal simulations for $\Delta T = 10~K$ and $\Delta T = 20~K$, orange dotted line and orange dashed line, respectively.}
  \label{fig.rrCurve}
\end{figure}

The PRF was maintained at 300~Hz in this work as a result of electronics limitations. We expect this can be improved, although we expect the thermal load on the chip and the build-up time of the cavity to limit the ultimate achievable PRFs. The analysis of build-up time has been previously considered \cite{kaspar_electro-optically_2012}. To further understand the PRF limitations the cavity build-up time was both measured and  compard to thermally-independent carrier dynamics and thermal simulations. The cavity build-up measurement was accomplished by triggering the Pockels cell twice during a single gain-switched pulse, where the first trigger allowed for any pump transients to be isolated then the build-up time of the cavity could be determined from the second pulse. The results of this measurement are shown in light blue circles in Fig.~\ref{fig.rrCurve}.

Simulated build-up time is determined by solving a set of coupled ordinary differential equations representing the carrier dynamics and a 2D cavity \cite{bedford_large_2018}. Parameters for this model match those given for the setup described above and a gain coefficient, $g_\circ$ = 2100~cm$^{-1}$, transparency carrier density, $N_{tr}$ = 2.0$\times$10$^{18}$~cm$^{-3}$, carrier transfer rates from the well to barrier (barrier to well), $\tau_{wb}$ = 20~ps ($\tau_{bw}$ = 200~ps), and total pump absorption of  $\Omega_w $ = 0.8.  Scattering loss is combined with the insertion loss of the Pockels cell and fit to be 2.26\%. The simulated, blue dashed line in Fig.~\ref{fig.rrCurve}, and measured build-up time match closely at low pump powers but the measured build-up time increases above $\approx$ 12~W of pump power leading to a decrease in the minimum PRF which we attribute to thermal effects that are not included in the rate-equation model in Ref.~\citenum{bedford_large_2018}. The inverse of the cavity build-up time would represent an upper limit of the achievable PRF at high-pump powers. The simulated results suggest a PRF of $\approx$ \(500~\si{\kilo\hertz}\) just below roll-over while the measurements suggest a limit closer to $\approx$ \(100~\si{\kilo\hertz}\). These values exceed the pump pulse width used currently though this could be reduced based on a shorter cavity length or a longer pump wavelength. 

For more insight, thermal simulations of the gain chip were performed using commercial finite element method (FEM) software \cite{COMSOL}. To determine the expected repetition rate of the cavity dumped VECSEL based on the heating of the VECSEL chip the full submount assembly was simulated with a top-hat 980~nm wavelength pump-beam incident on the front surface diamond heat spreader. The pulse length of 24~\si{\micro\second}\ is kept constant while the duty cycle and pump power are varied. To minimize the computational resources required the simulations are performed in 2D axisymmetric coordinates. Similar to the results reported previously \cite{holl2015recent}, the active region and DBR are treated as effective media with anisotropic thermal conductivities ($k_{gain}^{z} = 10.6$, $k_{gain}^{x,y} = 10.6$, $k_{DBR}^{z} = 14.3$, and $k_{DBR}^{x,y} = 20.1$ all in units of W/m$\cdot$K). Material parameters and nominal layer thicknesses are detailed above or taken from Ref.~\citenum{holl2015recent} and built-in parameters for the radiative beam in absorbing medium solver. To find the maximum temperature change relative to the single pulse or no-thermal memory regime, as the laser is operating at a PRF of \(300~\si{\hertz}\), a typical simulation of \(10~\si{\milli\second}\) is required with \(1~\si{\micro\second}\) interval to fit the saturation temperature. The resulting maximum temperature within the VECSEL chip is determined for each pulse and a saturation temperature is fit. The temperature difference between a single pulse and a higher duty cycle is then taken to be $\Delta T$ = 10~K (20~K) for a maximum acceptable duty cycle. The lower $\Delta T$ is indicative of more sensitive characteristics such as intracavity spectral filtering (e.g. alignment of the gain to the diamond heat-spreader) while the higher $\Delta T$ is more indicative of thermal impact to optical gain. For a constant pump pulse width, the duty cycle can be represented as a thermal relaxation time for the \(24~\si{\micro\second}\) pump pulse, represented as orange dotted (dashed) lines in Fig.~\ref{fig.rrCurve}. While utilizing an effective medium approach doesn't account for photon emission at the resonant wavelength or reflections of the pump laser the simulated increase in temperature is consistent between the measured spectral shift of the laser output and the spectral shift of 0.21 - 0.25~nm/K reported in similar VCSELs \cite{sanchez2012single} and VECSEL gain chips \cite{hopkins2007tunable}.  Typically, a $\Delta T$ of 10~K will have an acceptable difference in output, both output power and spectral shift, while a $\Delta T$ of 20~K may lead to an acceptable drop in output power in return for an increase in the PRF. The conservative estimate ensures there is minimal change in the gain of the active region as well as no appreciable shift in the DBR resonance other than at rollover. These results suggest the limiting factor for the PRF, $\approx $\(6.5~\si{\kilo\hertz}\) ($\approx $\(20~\si{\kilo\hertz}\)) when a 10~K (20~K) temperature change can be sustained, to be thermal relaxation in the gain chip needed for the high output powers.

It should be reiterated that the build-up time calculations do not include thermal aspects and the thermal modeling does not include the carrier recombination leading to upper and lower bounds of the expected PRF, respectively.  For a full estimation of the PRF these simulations would need to be coupled which is beyond the scope of the present work which provides a trend that is consistent with the increased turn-on time resulting from higher pump powers. Additionally, neither simulation takes into account the observed induced birefringence in the Pockels cell. This is the likely reason for the increase in the measured build-up time of the cavity at pump powers above $\approx$ 12~W. An increase in the PRF would compound this effect resulting in a higher cavity loss at higher pump powers. This could likely be mitigated by reducing the necessary pump pulse width either by decreasing the cavity volume or by pumping the VECSEL gain at longer wavelengths, such as 1470~nm for a lower thermal load \cite{holl2015recent}.

The semiconductor gain medium allows for designing the peak gain and emission of the laser to be in the spectral region at the the long wavelength tail of thulium and on the short wavelength side of holmium emission. This aspect could be useful for  applications including CO$_2$ sensing as well as in other applications where the output power and pulse characteristics are the defining figure of merit. While rare-earth lasers allow for high energy storage in a large gain volume, that can't be achieved in VECSELs; a direct comparison to the approaches for high peak powers in the two micron spectral range is limited to thin disk and other small gain medium capable of pulses in the nanosecond range. Several thulium and holmium works fit this criteria. Cavity dumping was utilized with a 16~mm long Tm:YAP (3.5\%) laser and an AOM to achieve peak power of 200 W with a PRF of 200 kHz and 43 ns pulses at a wavelength of 1990~nm \cite{yao2014diode}. Similar results, 194~W peak power, were found in a shorter but more highly doped crystal, 9~mm long Tm:YAG (4\%), with an AOM where 31~ns pulses{ were achieved \cite{dai2016output}.
To significantly increase the peak output power for a small gain volume rare-earth cavity dumped laser it must be used in combination with $Q$-switching. Utilizing this scheme with a 45 mm long Tm:LuAG (2\%), a 100~kHz 4.1~kW peak power output with 3 ns pulses was achieved \cite{zhang2016electro}. For holmium, cavity dumped results are limited but with a 40 mm long Ho:SSO (0.5\%) gain crystal and a Pockels cell 35~kW (17~kW) peak power was achieved at a PRF at of 50~kHz (100~kHz) while utilizing a volume Bragg grating to limit the linewidth to 70~pm at a wavelength of 2100~nm \cite{duan2018electro}. At the higher-end of the nanosecond pulse regime Q-switching using Cr$^{2+}$:ZnSe as the saturable absorber element with 4~mm thich Ho:SSO (1\%) has produced $\approx$ 100~ns pulses in the 2100~nm spectral range with a pulse energy of about a 125~\si{\micro \joule} \cite{yang2019passively}, or roughly 1.1~kW peak power, with spectrally pure outputs and linewidths below a nanometer. In each of these cases the gain volume is significantly larger than in a VECSEL, $>$4 orders of magnitude, but in terms of peak power the VECSEL output fits well between the purely cavity dumped thulium results and the cavity dumped $Q$-switch results in thulium and holmium systems. Additionally, the VECSEL results in a single transverse mode maintaining high beam quality, where the thulium and holmium results available in the literature show $M^2$ values in the 1.3 - 1.5 range, suggesting multiple transverse modes. A 0.5~kW peak power output from a compact single transverse mode semiconductor laser in the two micron spectral region capable of nanosecond pulses on-demand has applications in LiDAR, nonlinear supercontinuum generation, material processing, and gas sensing.

\section{Conclusion}\label{sec:5}

A \(\lambda = 2.04~\si{\micro\meter}\) cavity dumped VECSEL capable of achieving \(512~\si{\watt}\) peak power in \(10~\si{\nano\second}\) pulses was demonstrated, corresponding to a maximum pulse energy of 5.03~\si{\micro \joule}. The operation wavelength is in between the typical spectral range of thulium and holmium thin disk laser outputs, is at a wavelength attractive for LiDAR applications, and exceeds the peak powers reported for these systems. The minimization of cavity loss and the gain chip optimization, better thermal dissipation, and more efficient 980~\si{\nano\meter} pumping, coupled with gain switching allowed for an improvement of  peak power by a factor of 17 as compared to  previous state of the art GaSb based VESCEL cavity dumping \cite{kaspar_electro-optically_2012}. Additionally, utilizing a TFP as the output coupler allows for a high quality symmetric beam with a nearly unity $M^2$ in both dimensions. Maximum repetition rates and impact upon attainable pulse energies were theoretically analyzed. These peak powers can be extended to PRFs in kilohertz range with improved electronics.

\section*{Acknowledgment}

The authors thank Jussi-Pekka Penttinen at VEXLUM for technical discussions and Joshua Myers now at Altamira Technologies Corp. for technical discussions and early characterization of components.

\end{document}